\documentclass[sigconf,authorversion,nonacm,screen,timestamp,urlbreakonhyphens]{acmart}

\usepackage{amsmath}
\usepackage{xfrac}
\usepackage{graphics}
\usepackage{parskip}
\graphicspath{{./}{./figures/}{./screenshots/}}

\title{On the design of text editors}
\author{Nicolas P. Rougier}
\orcid{0000-0002-6972-589X}
\affiliation{
  \institution{Inria Bordeaux Sud-Ouest}
  \streetaddress{200 Avenue de la vieille tour}
  \city{Bordeaux}
  \postcode{33405}
   \country{France}
}
\email{Nicolas.Rougier@gmail.com}
\acmYear{2020}
\setcopyright{acmlicensed}
\acmConference[CHI 2021]{Computer Human Interaction}{May 8--13}
                         {Yokohama, Japan}



\begin{document}
\begin{teaserfigure}
  \begin{center}
    %% \makeatletter%
    %% \if@twocolumn%
      \includegraphics[height=5.605cm]{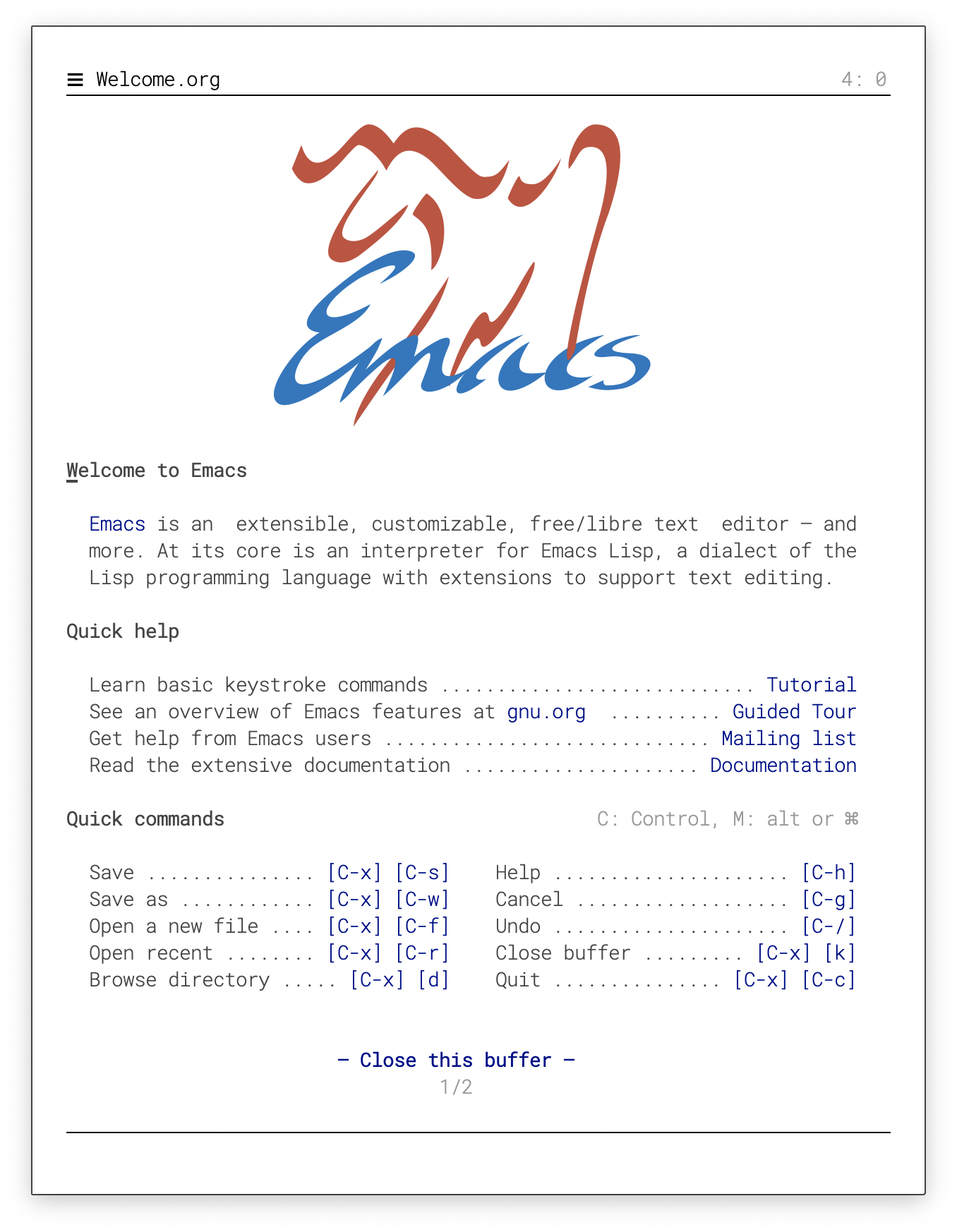}
      \includegraphics[height=5.605cm]{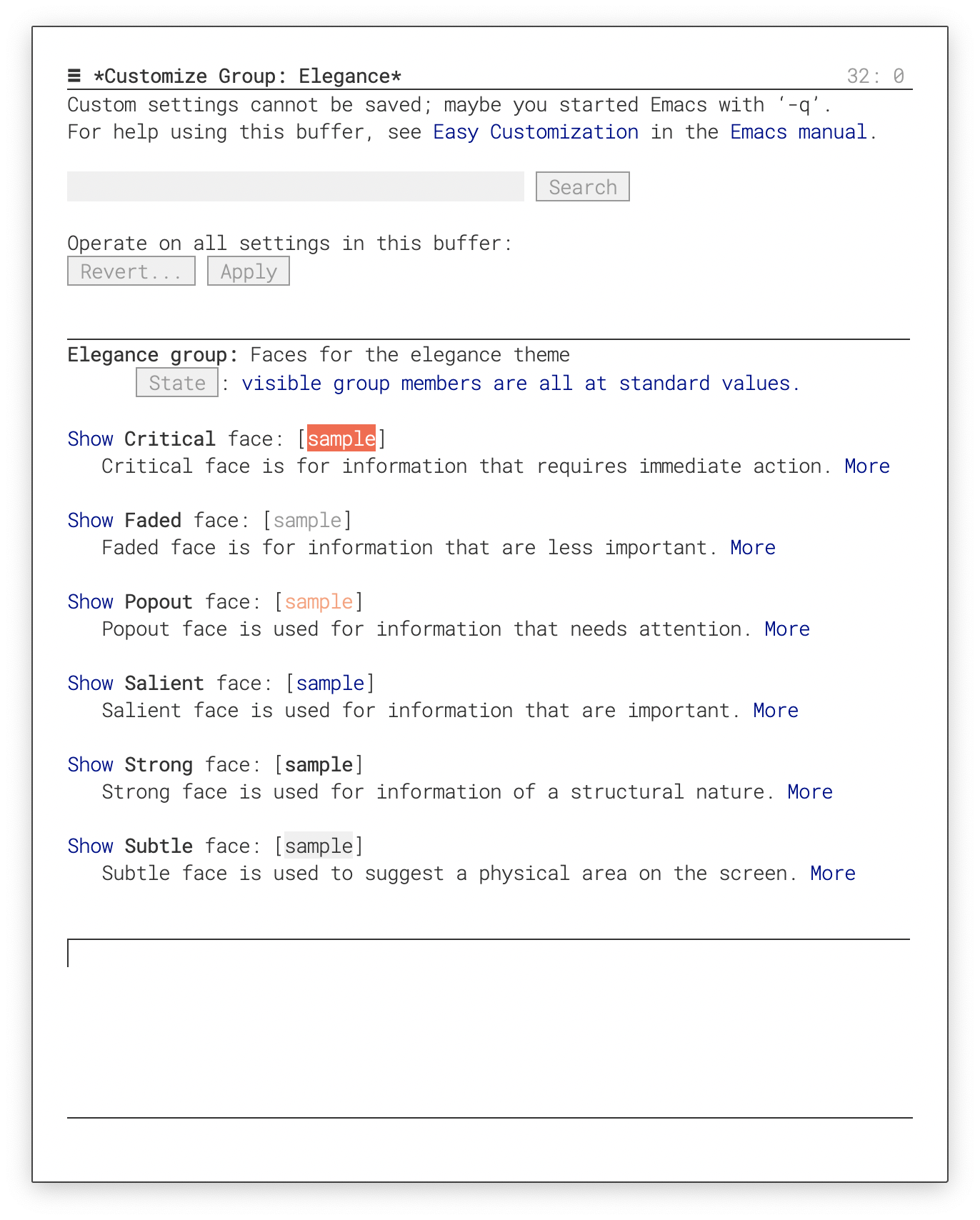}
      \includegraphics[height=5.605cm]{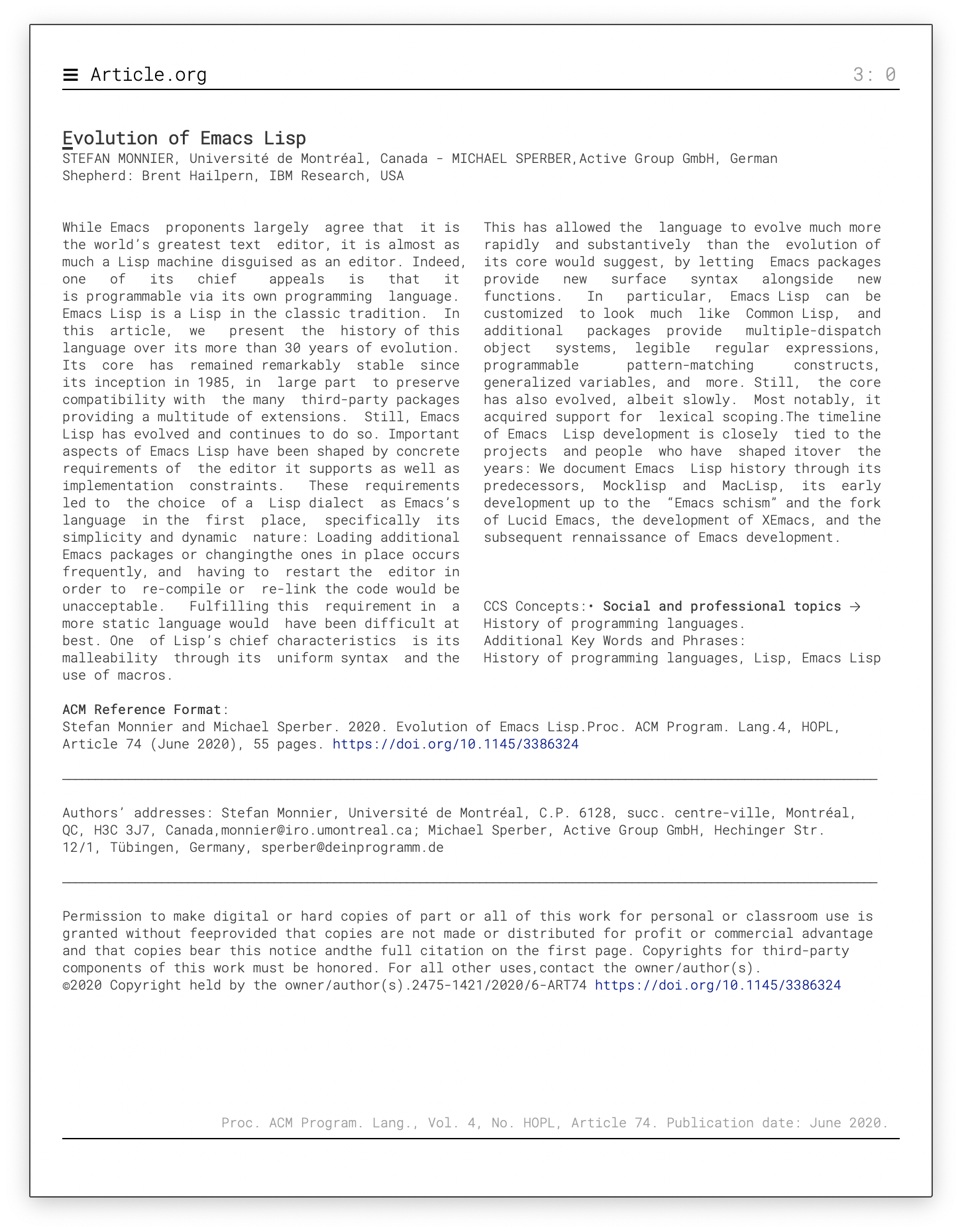}
      \includegraphics[height=5.605cm]{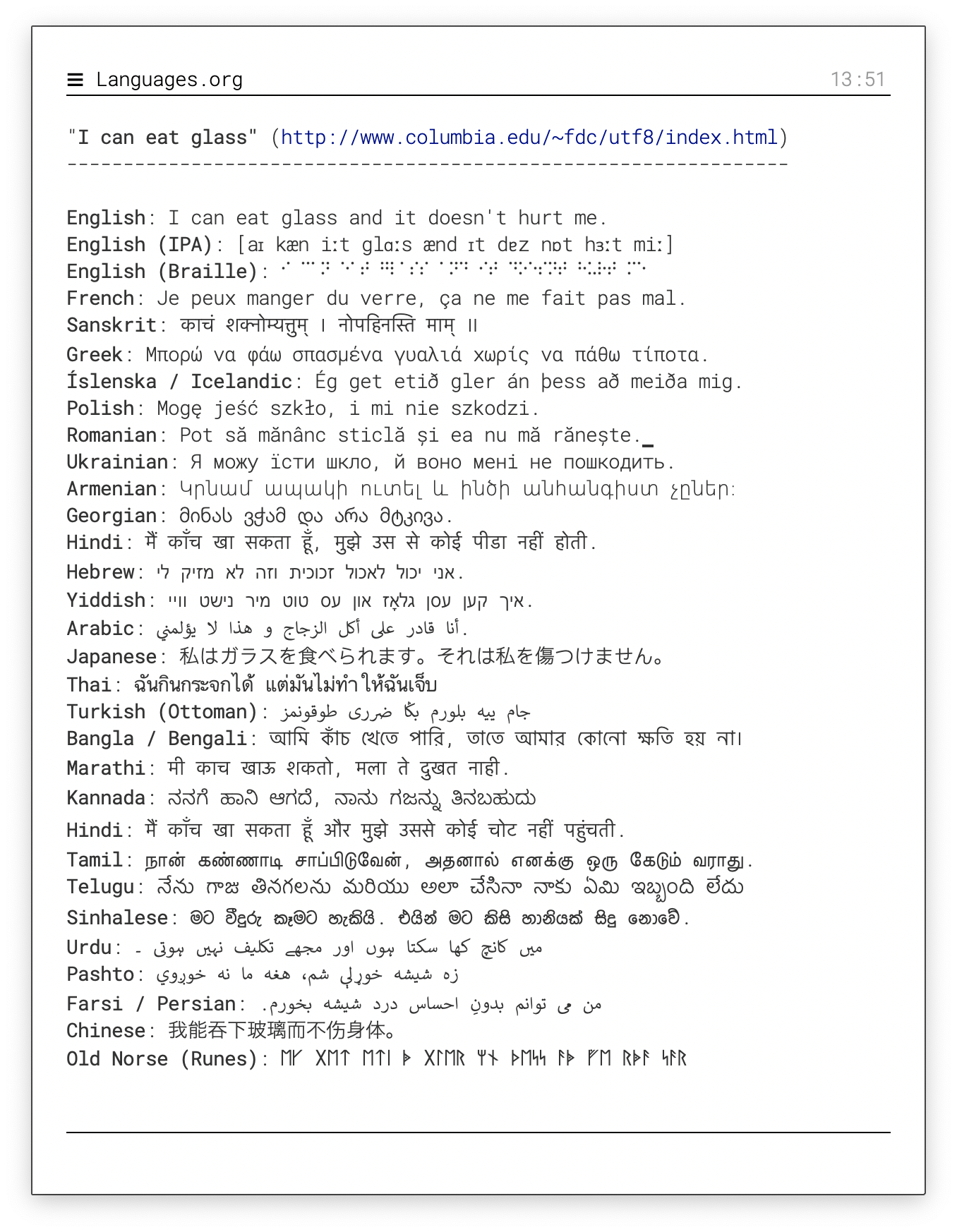}
    %% \else
    %%   \includegraphics[height=4.75cm]{teaser-1}
    %%   \includegraphics[height=4.75cm]{teaser-2}
    %%   \includegraphics[height=4.75cm]{teaser-3}
    %%   \includegraphics[height=4.75cm]{teaser-4}
    %% \fi
    %% \makeatother
  \end{center}
  \caption{GNU Emacs screenshots with hacked settings.}
  \Description{}
  \label{fig:teaser}
\end{teaserfigure}

\begin{abstract}
  Code editors are written by and for developers. They come with a large set of
  default and implicit choices in terms of layout, typography, colorization and
  interaction that hardly change from one editor to the other. It is not clear
  if these implicit choices derive from the ignorance of alternatives or if they
  derive from developers' habits, reproducing what they are used to. The goal of
  this article is to characterize these implicit choices and to illustrate what
  are alternatives, without prescribing one or the other.
\end{abstract}

\maketitle
\section{Introduction}

Text (or code) editors are computer programs used for editing plain text. The
first rudimentary text editor (QED) was released in 1965
\cite{Deutsch:1967,Ritchie:2004}, soon followed by EDIT
\cite{Bourne:1971}. Since then, a plethora of text editors have been written,
such as GNU Emacs \cite{Stallman:1981} (1976) and vi/vim \cite{Target:2018}
(1976), for the oldest (and still actively developed), or Atom (2014) and
Sublime Text (2008), for the more recent ones. Each editor offers some specific
features (e.g. modal editing, syntax colorization, code folding, plugins, etc.)
that is supposed to differentiate it from the others at time of release. There
exists however one common trait for virtually all of text editors: they are
written by and for developers. Consequently, design is generally not the primary
concern and the final product does not usually enforce best recommendations in
terms of appearance, interface design or even user interaction. The most
striking example is certainly the syntax colorization that seems to go against
every good design principles in a majority of text editors and the motto guiding
design could be summarized by {\em ``Let's add more colors''} (using regex).\\

More generally, modern text editors comes with a large set of default and
implicit choices that hardly change from one editor to the other. To take only
one example, most editors (that I know) consider that there exists only two font
weights ({\em regular} and {\em bold}): you can choose the {\em regular} font
but rarely can you choose the boldness of the {\em bold} one. Consequently, if
you choose a light or thin weight for the regular font, the difference with the
bold font is dramatically accentuated. It is not clear to me if these implicit
choices derive from the ignorance of alternatives or if they derive from
developers' habits, reproducing what they are used to. The goal of this article
is thus to characterized these implicit choices and to illustrate what are the
alternatives. However, I do not recommend any specific alternative since this
would require a user study that has yet to be done \cite{Roberts:1983}. The goal
is rather is to make the developer aware of the alternatives and to let her
experiment.

\textbf{Note:} In this landscape, GNU Emacs and vi (or VIM) are very specific
because they're highly hackable such that the advanced user can configure them
to their exact liking. The screenshots from the teaser figure have been designed
using GNU Emacs.

\section{Layout}

Beyond the actual text editing area where one can edit a file, most code editors
comes with a number of additional features such as tabs, status bar, console,
file browser, minimap, etc. This list may vary from one editor to the other, but
it appears that there is a set of minimal features that a modern text editor is
expected to offer.
\begin{figure}
  \includegraphics[width=.475\textwidth]{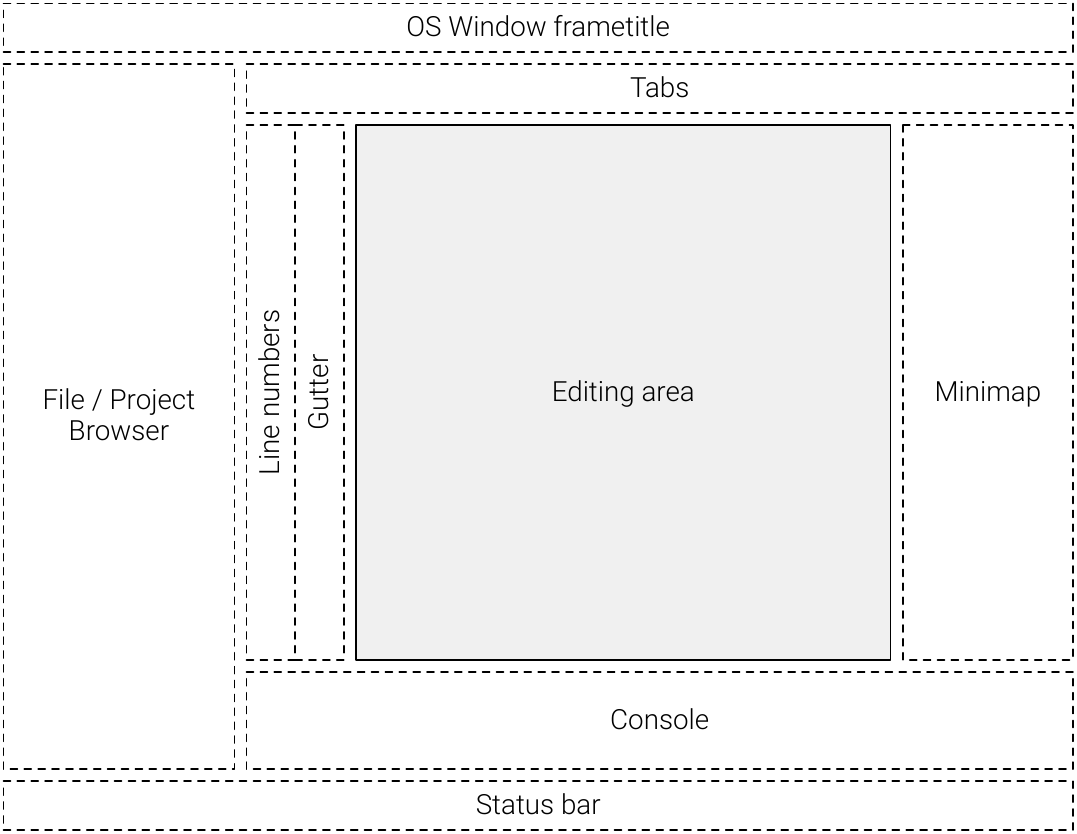}
  \caption{Default layout found in several modern editors}
  \Description{}
  \label{fig:layout}
\end{figure}
When these features are all active at once, this naturally and drastically
reduces the size of the main editing area as illustrated on figure
\ref{fig:layout}. More importantly, this clutters the space with secondary and
peripheral information that a user doesn't look at or use very often when
writing code. For example, the file (or project) browser is useful for giving an
overview of a project or to select a specific file. But a developer spends most
of her time in the text editing area and these peripheral information can be
largely considered as a distractor or lost space. This might be the reason why a
lot of editors allow to deactivate these components and some of them even offer
a distraction free or zen mode (natively or via plug-in) where most of the
peripheral information is actually hidden.

\begin{figure}[htbp]
  \includegraphics[width=.475\textwidth]{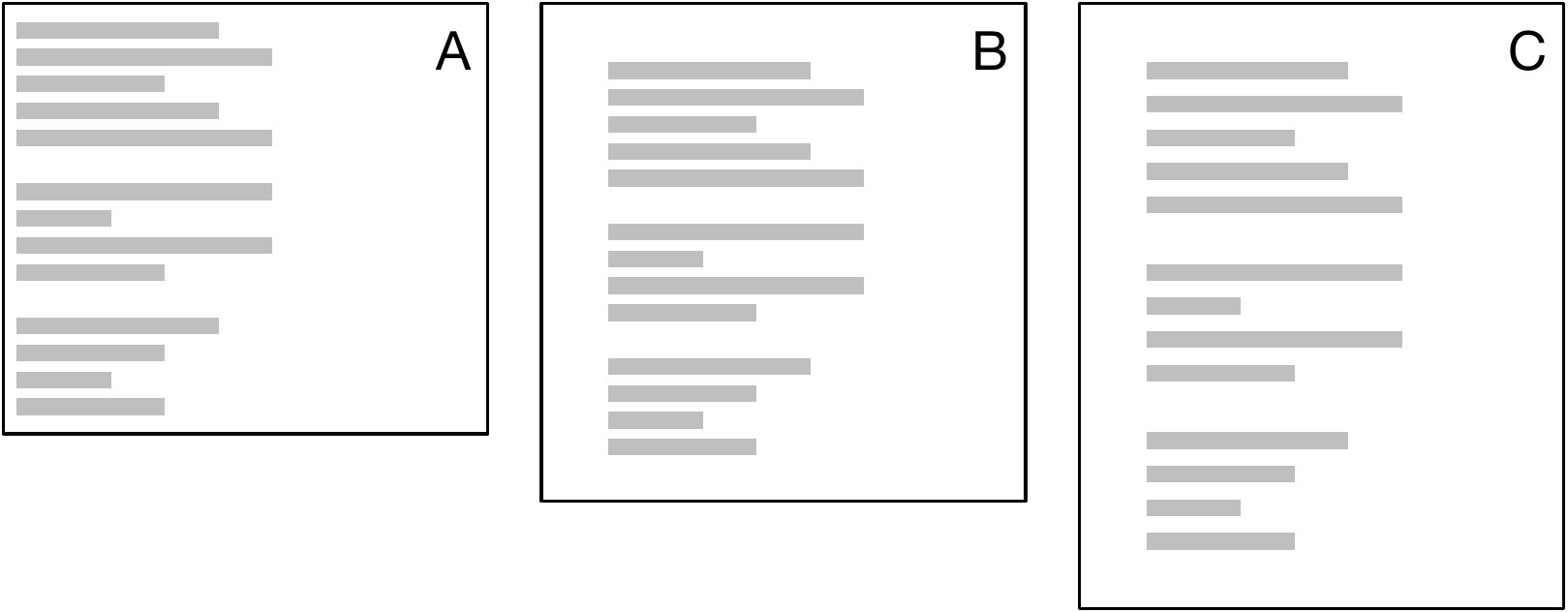}
  \caption{Influence of shape, margins and line spacing}
  \Description{}
  \label{fig:margins}
\end{figure}

If we now look closer at the single window mode (without any extra features), it
is striking to observe that there are generally hardly any margins as
illustrated on figure \ref{fig:margins}A. While it is not conceivable to have a
book or a PDF documents without margins, it is somehow considered perfectly
normal to not have margins for a code editor (and for a terminal as well on that
matter). It is even more surprising considering the recommendations that were
made as early as 1986 by \citet{vannes:1986}. Nonetheless, only a handful of
modern editors offers the possibility to define margins. One possible
explanation might be historical. Forty years ago, the standard vt100 terminal
offered only a resolution of 80$\times$24 characters and we could easily imagine
that with such limited display, margins were out of the question. Today (2020),
most screens have a much larger capability and it is thus difficult to
understand why, if not habits, margins are still not an option. This would ease
the reading (see \ref{fig:margins}B) and the same is true for line spacing that
is generally set to its minimal value without the possibility to modify it while
it would ease the reading even more.

Furthermore, it seems that the default shape of the window in single window mode
has inherited the $\sfrac{3}{1}$ ratio from the vt100 console and the default
window tends to be larger than tall. This is again a peculiar choice considering
that code is mostly made of short lines (best practices recommend lines length
between 72 and 90 characters, 80 being an heritage of IBM punch cards, see
\cite{Nanavati:2005} for a review) that are mostly written on the left on the
window. It would thus make sense to have taller windows as illustrated on figure
\ref{fig:margins}C. Of course, this shape can be modified from the window
manager or the settings, but there could be an option, for example, to enforce a
fixed ratio of $\sfrac{1}{\sqrt{2}}$ (ISO 216, see figure \ref{fig:teaser}).

\section{Typography}

Typography is the poor relation of code edition \cite{Oman:1990}. It is as if
typography recommendations had been frozen sometime during the eighties and nothing
has ever changed since then. Typography in most code editors can be summarized
as {\em Use only monospaced fonts with two weights (regular \& bold) and two
  slants (normal \& italic)}. However, digital typography has changed a lot
since the eighties \cite{Bigelow:2020a,Bigelow:2020b} and, for example, most
typefaces come in several weight variations, ranging from ultra thin to ultra
black such that it is possible to define several couples. For example, in the
case of Roboto Mono, we can use thin, light, regular, medium or bold:

\includegraphics[width=8cm]{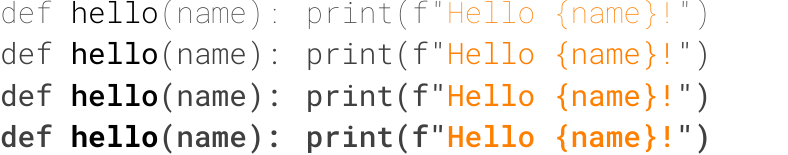}

%% {\thin \textcolor{darkgray}{def
%%     {\light \textcolor{black}{hello}}(name):
%%     print(f"\textcolor{orange}{Hello \{name\}!}")}}\\
%% {\light \textcolor{darkgray}{def
%%     {\regular \textcolor{black}{hello}}(name):
%%     print(f"\textcolor{orange}{Hello \{name\}!}")}}\\
%% {\regular \textcolor{darkgray}{def
%%     {\medium \textcolor{black}{hello}}(name):
%%     print(f"\textcolor{orange}{Hello \{name\}!}")}}\\
%% {\medium \textcolor{darkgray}{def
%%     {\black \textcolor{black}{hello}}(name):
%%     print(f"\textcolor{orange}{Hello \{name\}!}")}}

Furthermore, with the advent variable fonts, we can consider several
alternatives. Variable fonts have been introduced in version 1.8 of the OpenType
font format specification. This new type of font includes (in a single file)
multiple variations over one or several axis such as weight, width, optical
size, slant etc and makes it possible to interpolate between these
variations. For the end-user, this means she can precisely define the different
typeface she wants to use. For a text editor, this means it is possible to have
context dependent and dynamic font variations. For example, it is possible to
have subtle weight variations of a text depending whether text is light text
over a dark background or the opposite (think about selected text). Since any
font axis can be made variable this offers a tremendous amount of possibility
and probably a tremendous amount of ways to abuse it.
%% \begin{figure}
%%   \includegraphics[width=.475\textwidth]{variable-font}
%%   \caption{Gingham variable font with continuous variation along width
%%     and weight axes.}
%%   \Description{}
%%   \label{fig:variable}
%% \end{figure}

Another typographical features that was hardly used until very recently are
ligatures, that is, the union of two or more glyphs into a single glyph. The
Hasklig font by Ian Tuomi (based on Source Code Pro) is the first font to have
taken advantage of ligatures and adapted them to source code. For example, the
usual notation for {\em greater than or equal} in most programming languages is
{\ttfamily >=} while the mathematical notation is $\geq$. Ligatures can be used
to actually display the mathematical notation without changing the source
code. Today font families such as Fira Code, Monoid, Iosevska, Inconsolata or
JetBrains Mono all offers a various amount of such code-oriented ligatures
\cite{Latin:2020}. Even if this feature does not entirely depend on the editor
(editor must enforce ligature and the font must possess ligatures), it is an
aspect to be considered when selecting the default font that is shipped with the
editor. Note however that some typographers do not recommend the usage of such
code ligature. \citet{Butterick:2013} goes a bit further and explains that
ligatures in programming fonts are a terrible idea because i) they contradict
unicode and ii) they are guaranteed to be wrong sometimes.

\begin{figure}
  \includegraphics[width=.475\textwidth]{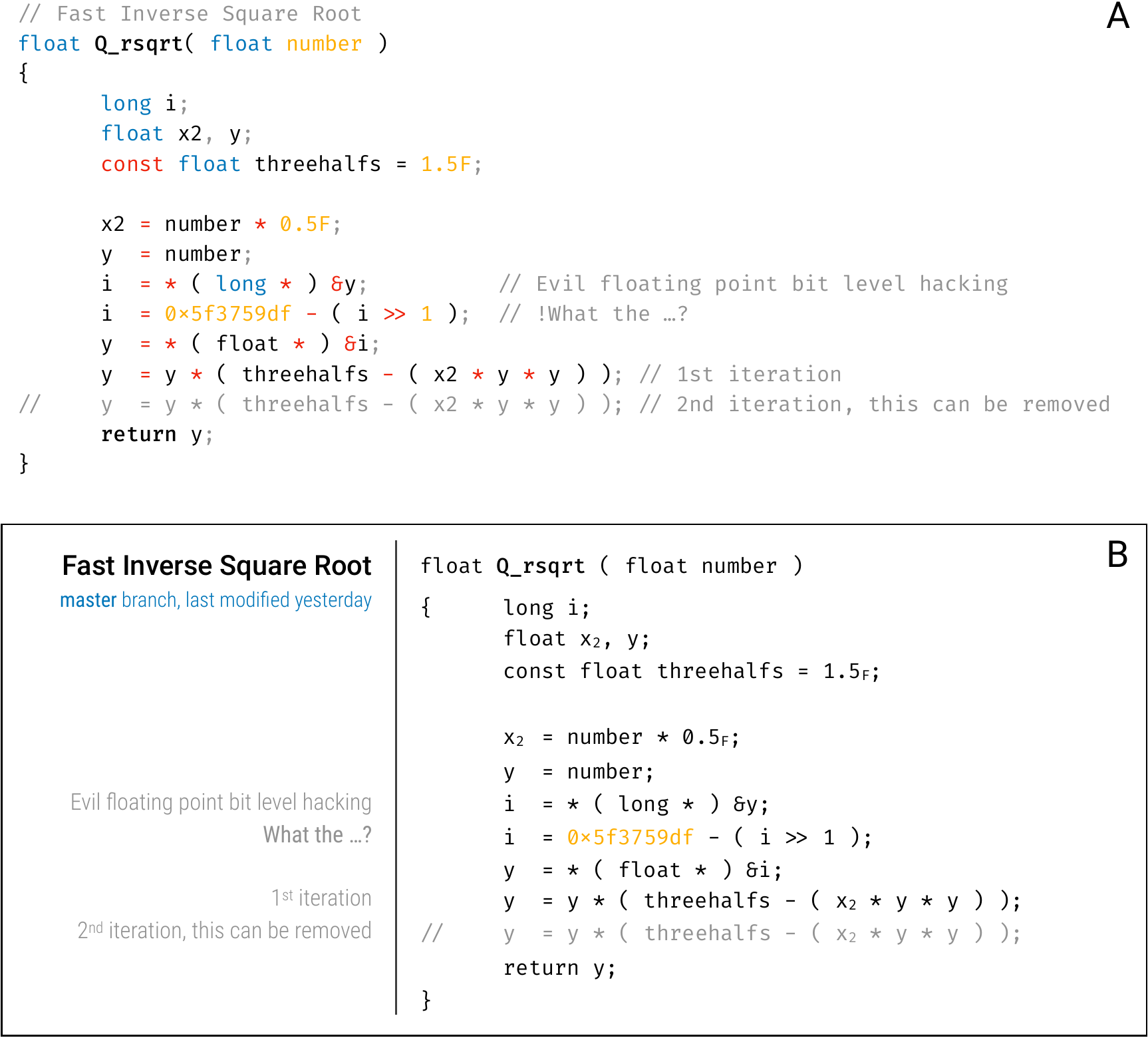}
  \caption{A. Regular code display using syntax highlighting. B. Alternative
    code display taking advantage of typography.}
  \Description{}
  \label{fig:mockup}
\end{figure}

Beyond eye-candy features, typography can also be exploited to introduce some
radical changes in code layout as shown on figure~\ref{fig:mockup}. In this
example, code and comments are separated in two distinct columns using two
different fonts (monospaced font (Fira Code) for code and condensed font (Roboto
Condensed) for comments). Comments are placed on the left and the short title is
made prominent using larger and heavier font, enriched with contextual
information (branch and last commit). It is interesting to note that this
alternative occupies the exact same physical space as a regular display. Of
course, this mockup is only one possibilities among several alternatives and
most probably, there are better ways to highlight what is deemed to be the
important information in a source code. But I think this simple example
illustrates the fact that typography can dramatically enhanced readability.

\section{Colorization}

Syntax highlighting is pervasive and entrenched in virtually all code editors.
It aims at visually distinguishing elements of the source code using different
colors and/or typefaces. The origin can be traced back to the patent filed in
1982 by \citet{Klock:1982} about a syntax error correction method and apparatus
(the first color terminal dates back to 1979 \cite{IBM:1979}).  This has been
further refined in 2009 with the concept of semantic highlighting
\cite{Nolden:2009} that uses knowledge of the underlying language to provide a
finer control on how a given element should be highlighted. However, the
advantages of syntax highlighting are far from being obvious. The most recent
study \cite{Hannebauer:2018} {\em could not find evidence in data that syntax
  highlighting as used in Eclipse has a beneficial effect on program
  comprehension for programming novices}. To make things worse, it is also not
rare to have a Christmas tree effect resulting from an abuse of syntax
highlighting \cite{Akesson:2007}. Consider for example figure \ref{fig:mockup}A
that displays a code snippet using the default syntax highlighting of a recent
editor. Even though the code is rather small, syntax highlighting results in six
different colors being used all over the source code, making it difficult to
assign a specific semantic to any given color. The question, is thus, does it
help the developer? According to \cite{Asenov:2016}, {\em using more visual
  variety when rendering methods substantially reduces comprehension time of
  code features}. The problem with syntax highlighting is that it does not seem
to be based on any specific principles and derives from the possibility of
identifying code parts based on regular expressions, and the colorization of
such expression. But there are no scientific recommendation on what to highlight
or how to highlight. Only the {\em solarized} color palette crafted by Ethan
Schoonover seems to enforce some design principles with {\em reduced brightness
  contrast while retaining contrasting hues for syntax highlighting readability}
even though the author doesn't prescribe how to apply it such color scheme.

There exist however alternate use of colorization where the semantic of color is
well defined. This is the case for multi-authored document where each author is
identified with a unique color. This can be used during live editing such as
notepads or post-edition using tools such as git blame (for example). Another
possibility is to use colorization in order to show the modification history of
a document, using light tint for old modifications and heavier tint for recent
modifications. \citet{Wayne:2020} goes a step further and denounces the use of
syntax highlighting since it is a waste of an important information channel and
suggest several alternative uses of color, among which, rainbow parenthesis,
context highlighting, import highlighting, argument highlighting, type
highlighting, etc.  Instead of syntax or semantic colorization based on content,
a simple alternative would be to adapt colorization to the reader, taking
attentional constraints into account \cite{Treisman:1980,McCayPeet:2012}. For
example, here is the color scheme that has been used to design the mockup on
figure \ref{fig:teaser} and which is based on the perception rather than the
content:

\begin{description}
\item[Critical] face is for information that requires immediate action. It
  should be of high contrast when compared to other faces. This can be done
  (for example) by setting an intense background color, typically a shade of
  red. It must be used scarcely.
  
\item[Popout] face is used for information that needs attention. To achieve such
  effect, the hue of the face has to be sufficiently different from other faces
  such that it attracts attention through the popout effect.

\item[Strong] face is used for information of a structural nature. It has to be
  the same color as the default color and only the weight differs by one level
  (e.g., light/regular or regular/bold). It is generally used for titles,
  keywords, directory, etc.

\item[Salient] face is used for information that are important. To suggest the
  information is of the same nature but important, the face uses a different hue
  with approximately the same intensity as the default face. This is typically
  used for links.

\item[Faded] face is for secondary information that is less important. It is
  made by using the same hue as the default but with a lesser intensity than the
  default. It can be used for comments, secondary information and also replace
  italic (which is generally abused anyway).

\item[Subtle] face is used to suggest a physical area on the screen. It is
  important to not disturb too strongly the reading of information and this can
  be made by setting a very light background color that is barely perceptible.
\end{description}

In fact, such cognitive colorization does not require any change in syntax in
highlighting engines. It only requires a restricted set of colors and a careful
selection of what information needs to be salient, faded or strong.

\begin{figure}
  \includegraphics[width=.475\textwidth]{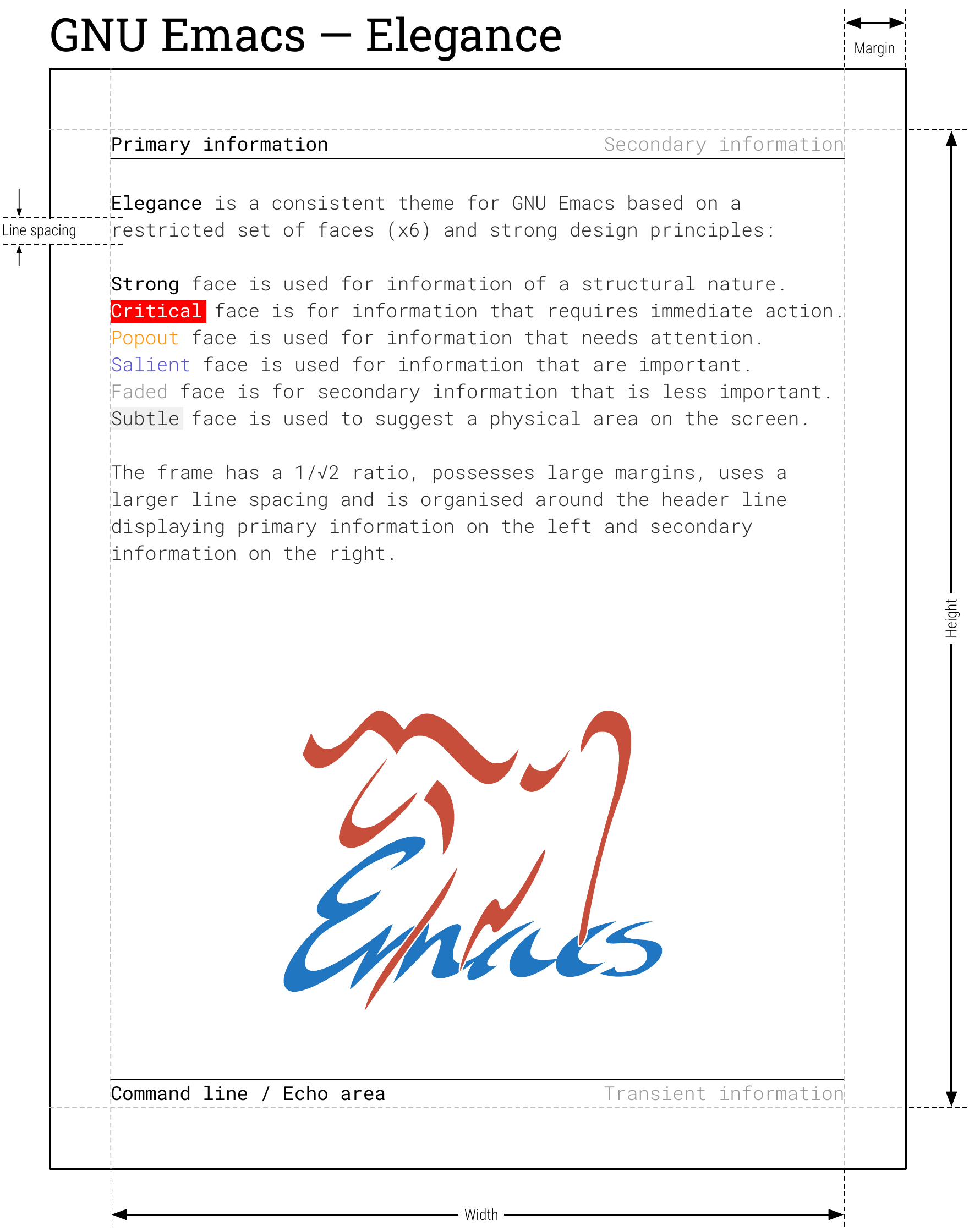}
  \caption{GNU Emacs with heavily modified settings enforcing some of the
    alternatives presented in this article.}
  \Description{}
  \label{fig:elegance}
\end{figure}

\section{User interaction}

One feature shared by all text editors is the linear representation of a text
file and the use of scrolling to navigate forward or backward. This has become a
seemingly ubiquitous part of our user experience and it seems difficult to
imagine any alternative that nonetheless exist. It is, for example, quite common
to have a dedicated navigation panel allowing to jump to a specific part of the
code. This panel can be built automatically (e.g. function, method, class,
definition) or manually using bookmark inside a text file. However, these
navigation panels usually follows the linear structure of the code (even though
some may propose a sorted list) and do not process semantically the content
(when built automatically). These navigation panels are actually comparable to a
table of content and this suggests a code source could be considered as a book
as it has been proposed by \citet{Oman:1990}. The author proposed to format
source code as if it was printed on a book, taking advantage of typography,
logical blocks separation and using the page paradigm (instead of scroll). As
explained by the author, {\em the components of a book (preface, table of
  contents, indices and pagination, chapters, sections, paragraphs, sentences,
  punctuation, type style, and character case) are all designed to facilitate
  rapid information access and transfer}. Such printed book paradigm has been
hardly used in any text editor but it is certainly a direction to explore
further.

An important and critical aspect of interaction in a code editor is (of course)
the actual input of text and commands since coding activity encompasses actual
code writing but also code navigation. In most modern editors, such inputs are
carried out via the combination of keyboard and mouse while, in older editors
such as Emacs and vi, it is possible to issue text and commands from keyboard
only. For Emacs, this is made possible using the command-line that is an
integral part of Emacs and allow to type literal commands. This has probably
inspired the command palette that is now found in modern editors. This command
line is complemented by several keybindings that may be global or specific to a
given mode (e.g. Python mode, Lisp mode, etc.) and any command can also be bound
to any key sequence (even very complex and long ones). The philosophy of vi is
quite different because it has adopted a modal approach and operates in two
modes: the insert mode where keystrokes modify the document and the command mode
where keystrokes are interpreted as commands. Such modal feature has been
somehow killed by \citet{Tesler:2012} who transformed the modal Bravo editor
into the modeless Gypsy editor during the seventies and whose efficiency has
been demonstrated \cite{Poller:1983}. This may very well had a profound impact
on the community because the only surviving modal editor is actually vi and only
the new kakoune editor (\href{https://kakoune.org/}{kakoune.org}) adopted a
similar modal approach (using a object + verb approach while vi uses verb +
object). However, four decades later, this efficiency might need to be
re-evaluated, especially in light of challenges such as vimgolf
(\href{https://www.vimgolf.com/}{www.vimgolf.com}) that demonstrates how to
perform complex and real-world code transformation with only a few
keystrokes. For example, let's consider the following text: ``The quick brown
fox jumps over the lazy dog.''  that we want to convert to ``The quick lazy dog
jumps over the brown fox.'' The actual sequence of vi keystrokes to transform
the first text into the second is \texttt{2wd2w3wPd3w6bep} (\texttt{2w}: Move
forward two words -- \texttt{d2w}: Delete two words -- \texttt{3w}: Move forward
three words -- \texttt{P}: Paste the previously deleted text before cursor
position -- \texttt{d3w}: Delete three words -- \texttt{6b}: Move back six words
-- \texttt{e}: Move to the end of the current word -- \texttt{p}: Paste the
previously deleted text after the cursor position). This keystroke sequence is
of course a bit cryptic for those not familiar with vi but it is nonetheless
quite efficient. No doubt that the casual user would not like to have to learn
such meta-language before being able to use the editor.
  
But let me remind you that users of code editors are not casual users and such
modal interface might need to be further exploited in modern editors. Similarly,
there are plenty of ``old'' concepts that might be worth to be re-considered
such as kill ring (extended clipboard), recursive undo (possibility to undo an
undo command), rectangle selection, etc.

\makeatletter
\if@twocolumn \balance \fi
\makeatother

\section{Conclusion}

I've highlighted several implicit choices that are present in a number of both
old and modern text editors and introduced several alternatives that, I think,
are worth to be explored and exploited by future developers and designers.
Several of these alternatives have been actually introduced 30 years ago by
\citet{Oman:1990} and I've mostly updated them in light of available
technologies in 2020 and added new ones that were hardly imaginable in 1990. In
the meantime, there are viable alternatives that are already implemented in
historic editors (vi and GNU Emacs) that may be also worth to be reconsidered in
the design of future code editors. Overall, the sum of all these alternatives
offer great possibilities to enhance the user experience as illustrated in
figure \ref{fig:elegance}. This is not a mockup but my actual GNU Emacs
configuration that I use for code, text, mail and agenda. Note that I'm not
advocating for people to use such setup, the goal is merely to highlight an
example of a radical design for a text editor that is perfectly usable according
to my daily and personal experience.

\bibliographystyle{ACM-Reference-Format}
\bibliography{draft.bib}

\end{document}